\documentclass[12pt,a4paper]{article}
\usepackage{jheppub}
\usepackage{chemarrow}
\usepackage{amsmath,amssymb,amsfonts,mathtools} 
\usepackage{float}
\usepackage{subfig}
\usepackage{scalefnt}
\usepackage{wrapfig}
\usepackage{fancybox}
\usepackage{ifsym,yfonts}
\newcommand{\bea}{\begin{eqnarray}\displaystyle}
\newcommand{\eea}{\end{eqnarray}}
\newcommand{\nn}{\nonumber}
\newlength{\arrow}
\newcommand{\figref}[1]{Fig.~\protect\ref{#1}}

{\setlength{\fboxsep}{15pt}
\setlength{\mylength}{\linewidth}%
\addtolength{\mylength}{-2\fboxsep}%
\addtolength{\mylength}{-2\fboxrule}%
\Sbox
\minipage{\mylength}%
\setlength{\abovedisplayskip}{0pt}%
\setlength{\belowdisplayskip}{0pt}%
\equation}%
{\endequation\endminipage\endSbox
\[\fbox{\TheSbox}\]}

\begin{document}
\title{Topological Field Theory Amplitudes for $A_{N-1}$ Fibration}
\author[a,b,c]{Amer Iqbal,}
\author[a]{Ahsan Z. Khan}
\author[a]{Babar A. Qureshi}
\author[d]{Khurram Shabbir}
\author[a]{Muhammad A. Shehper}
\affiliation[a]{Department of Physics, School of Science \& Engineering, LUMS, D.H.A, Lahore, Pakistan}
\affiliation[b]{Department of Mathematics, School of Science \& Engineering, LUMS, D.H.A, Lahore, Pakistan}
\affiliation[c]{Abdus Salam School of Mathematical Sciences, Government College University, Lahore, Pakistan}
\affiliation[d]{ Department of Mathematics, Government College University, Lahore, Pakistan}

\abstract{
We study  the partition function ${\cal N}=1$ 5D $U(N)$ gauge theory with $g$ adjoint hypermultiplets and show that for massless adjoint hypermultiplets it is equal to the partition function of a two dimensional topological field on a genus $g$ Riemann surface. We describe the topological field theory by its amplitudes associated with cap, propagator and pair of pants. These basic amplitudes are open topological string amplitudes associated with certain Calabi-Yau threefolds in the presence of Lagrangian branes. }

\maketitle

\section{Introduction}
The engineering of supersymmetric gauge theories in string theory has provided quite an amazing web of dualities connecting these theories. In recent years some supersymmetric indices associated with supersymmetric gauge theories have been related to two dimensional topological field theories \cite{Gadde:2009kb, Gadde:2010te, Gadde:2011ik, Gadde:2011uv,Fukuda:2012jr,Tachikawa:2012wi}. It was shown in \cite{BP, Aganagic:2004js} that the topological string partition function of a Calabi-Yau threefold which is a $\mathbb{C}^2$ bundle over genus $g$ Riemann surface in a certain limit can be obtained from a two dimensional topological field theory on the genus $g$ Riemann surface.  It was shown in \cite{Aganagic:2004js} that the topological field theory amplitudes are given by open string amplitudes associated with certain Calabi-Yau threefolds and that the underlying two dimensional topological field theory is q-deformed Yang-Mills.

The Calabi-Yau threefold which is a $\mathbb{C}^2$ fibration over genus $g$ Riemann surface gives rise to, via geometric engineering, $U(1)$ gauge theory with $g$ adjoint hypermultiplets.
In this paper we generalize this picture to the case of $U(N)$ gauge theory with $g$ adjoint hypermultiplets and show that the supersymmetric partition function of this gauge theory can be built using two dimensional topological field theory rules from certain amplitudes. Just like in the $U(1)$ case these amplitudes, which are the building blocks of the gauge theory partition function, are the open topological string amplitudes associated with certain Calabi-Yau threefolds. These Calabi-Yau threefolds are pieces of the Calabi-Yau threefolds which is a fibration over a genus $g$ curve with fibers which are resolution of $A_{N-1}$ singularity. By splitting the base genus $g$ curve in to cap, propagator and pair of pants we get a corresponding decomposition of the Calabi-Yau threefold into pieces which are $A_{N-1}$ fibration over the cap, propagator and the pair of pants. The open string amplitudes obtained by placing branes on the boundaries of these pieces give the corresponding TFT like amplitudes. The two dimensional TFT to which these amplitudes belong is just the quiver q-deformed Yang-Mills. The topological string partition function for the $A_{1}$ fibration over a genus $g$ curve was discussed in \cite{Diaconescu:2005ik} and a conjecture for the $N>2$ was made.  Our results are consistent with theirs for the unrefined case.

The paper is organized as follows. In section two we discuss the partition function of the $U(N)$ gauge theory with $g$ adjoint hypermultiplets. In section three we review the case of the TFT amplitudes for the $U(1)$ case. In section 4 we discuss the TFT amplitudes for the $U(N)$ case but for the omega background with $\epsilon_{1}+\epsilon_2=0$ i.e., the unrefined case. In section 5 we generalize results of section 4 and discuss the refined TFT amplitudes. Notation and definitions needed for some formulas are given in appendix A.

\section{5D Gauge Theory with $g$ Adjoint Hypermultiplets}

The 5D gauge theory with $U(N)$ gauge group and $g$ adjoint hypermultiplets can be geometrically engineered by taking M-theory compactified on a Calabi-Yau threefold which is an $A_{N-1}$ singularity fibred over a genus $g$ curve \cite{Katz:1996ht,Witten:1996qb}.  The partition function of this theory is given by an equivariant index on the instanton modulis space and can be obtained using Nekrasov's instanton calculus. The case of $N=1$ was discussed in \cite{Chuang:2012dv} and we generalize their result for $N>1$. Let us denote by $M_{N,k}$ the $U(N)$ instanton moduli space of charge $k$. $M_{N,K}$ is given by a hyper-K\"aler quotient with $\text{dim}_{\mathbb{R}}\big(M(N,k)\big) = 4Nk$ and is defined through the following quotient:
\bea
{\cal M}_{N,k}/U(k)
\eea
where 
\bea\nn
{\cal M}_{N,k}=\{(B_{1},B_{2},i,j)~|~[B_1,B_2]  + ij =0\,,
[B_1 , B_1^{\dagger}] + [B_2, B_2^{\dagger}] - ii^{\dagger} - jj^{\dagger} = \zeta Id\} 
\eea
and $B_{1,2}\in \text{End}(\mathbb{C}^k), i \in \text{Hom}(\mathbb{C}^k, \mathbb{C}^{N}),  j \in \text{Hom}(\mathbb{C}^N, \mathbb{C}^{k})$. The $U(k)$ action is defined as
\bea g(B_1, B_2, i, j) = (g B_1  g^{-1}, g B_2 g^{-1}, g i, j g^{-1}). 
\eea 
$M(N,k)$ has a $U(1)^N \times U(1)_{\epsilon_1} \times U(1)_{\epsilon_2}$ action defined on it:
\bea \label{action}
(B_1,B_2,i,j)\mapsto (e^{i\epsilon_1} B_{1},e^{i\epsilon_2} B_{2},i\,e^{-1}, e\,j)\,,
\eea
where $e=\mbox{diag}(e_1,e_2,\mathellipsis,e_N)$. As was shown in \cite{Nakajima} the fixed points of this action are in one-to-one correspondence with $N$-tuples of partitions $(\nu_1, \dots, \nu_N)$ such that $|\nu_1| + \dots |\nu_N| = k$, where $|\nu|$ denotes the sum of the parts of the partition $\nu$.  At a fixed point labelled by $(\nu_{1},\nu_{2},\cdots,\nu_{N})$ the weights of the $U(1)^{N+2}$ action on the tangent bundle, denotes by $w_{a}$, are given by 
\bea\label{tangentweights}
\sum_{a}e^{w_{a}}=\sum_{\alpha,\beta=1}^{N}e_{\beta}e_{\alpha}^{-1} \Big(\sum_{(i,j)\in \nu_{\alpha}}q^{-(\nu_{\beta,j}^{t}-i)}t^{-(\nu_{\alpha,i}-j+1)}+\sum_{(i,j)\in \nu_{\beta}}q^{\nu_{\alpha,j}^{t}-i+1}t^{\nu_{\beta,i}-j}\Big).
\eea

Denote by $E^{g}_{N,k}$ the following vector bundle on $M_{N,k}$:
\bea
E^{g}_{N,k} = {
\cal V}_{g}  \otimes {\cal L}_{g,p}\,.
\eea
In the above equation ${\cal V}_{g}=\big(T^* M_{N,k} \big)^{\oplus g}$ is the direct sum of $g$ copies of the cotangent bundle, ${\cal L}_{g,p}=\text{det}(W_{N,k})^{\otimes(1-g-p)}$ where $W_{N,k}$ is the tautological bundle over $M(N,k)$. The bundle ${\cal L}_{g,p}$ introduces the Chern-Simons coefficient in the partition function. The partition function of the $U(N)$ gauge theory with $g$ adjoint hypermultiplets is then given by:
\bea \label{PF1} Z^{g}_{N}&=&\sum_{k \geq 0} Q_{\bullet}^{k} Z^{g}_{N,k}\,,\\\nonumber
Z^{g}_{N,k}&=&\int_{M_{N,k}}\mbox{Ch}({\cal L}_{g,p})\mbox{Ch}_{\vec{y}}({\cal V}_{g})\mbox{Td}(M_{N,k}).
\eea
Where $\text{Td}(X)$ is the Todd class of $X$ and $\text{Ch}_{\vec{y}}(V) = \prod_{a=1}^{g}\prod_{i=1}^{s} (1-y_{a} e^{-\tilde{x}_i})$ is defined in terms of the $\widetilde{x}_i$, the Chern roots of $V$. Integral over $M_{N,k}$ is calculated using equivariant action of $T^{N+2}$ on $M_{N,k}$. $Z^{g}_{N,k}$ gets contribution from the fixed points of $T^{N+2}$ action of $M_{N,k}$ which are labelled by $N$-tuple of partitions $\{\nu_{1},\cdots,\nu_{N}\}$,
\bea\label{fixedpoint}
Z^{g}_{N,k}=\sum_{|\vec{\nu}|=k}Z^{g}_{N,\vec{\nu}}\,,\,\,\,\,Z^{g}_{N,\vec{\nu}}=f_{\vec{\nu}}^{p+g-1} \prod_{\alpha,\beta=1}^{N}\frac{\prod_{a=1}^{g}N_{\nu_{\alpha}\nu_{\beta}}(y_{a}Q_{\beta\alpha})}{N_{\nu_{\alpha}\nu_{\beta}}(y_{a}Q_{\beta\alpha})}
\eea
where $N_{\lambda\mu}(x)$ is the Nekrasov factor defined as
\bea\label{NekrasovFactor}
N_{\lambda\mu}(x)=\prod_{(i,j)\in \lambda}(1-
x\,q^{-\mu_{j}^{t}+i}t^{-\lambda_{i}+j-1})\times\prod_{(i,j)\in \mu}(1-x\,q^{\lambda_{j}^{t}-
i+1}t^{\mu_{i}-j})
\eea
and $\{Q_{\alpha \beta},y_{a}\} =\{e_{\alpha}e_{\beta}^{-1}, e^{2\pi i m_{a}}\}$. $\{m_{a}~|~a=1,\cdots,g\}$ are the masses of the adjoint hypermultiplets. We have defined parameters $q$ and $t$ such that $(q,t)=(e^{i\epsilon_{1}}, e^{-i\epsilon_{2}})$. $\{e_{\alpha}, q, t\}$ then are the parameters associated with the $T^{N+2}$ equivariant action on $M_{N,k}$. The factor $f_{\vec{\nu}}^{p+g-1}$ is the weight of the line bundle ${\cal L}_{g,p}$ as is given by,
\bea
f^{p+g-1}_{\vec{\nu}}:=\prod_{\alpha=1}^{N} \big(e_{\alpha}^{-|\nu_{\alpha}|}t^{\frac{||\nu^t||^2}{2}}q^{-\frac{||\nu||^2}{2}}\big)^{p+g-1}\,.
\eea
It is a product of factors corresponding to each of the Cartan $U(1)$'s given in \cite{Chuang:2012dv}.

After some simplification the partition function given in Eq.(\ref{PF1}) can be written as
\bea\label{PF2}
Z^{g}_{N}({\bf m},\epsilon_1,\epsilon_2)&=&
\sum_{\vec{\nu}}\widetilde{Q}^{|\vec{\nu}|}f_{\vec{\nu}}^{p+g-1}\frac{\prod_{a=1}^{g}L_{\vec{\nu}}(y_{a})R_{\vec{\nu}}(y_{a})}{L_{\vec{\nu}}(1)R_{\vec{\nu}}(1)}
\eea
where
\bea\nn
\widetilde{Q}&=&Q_{\bullet}(-1)^{(g-1)N}\Big(\prod_{a=1}^{g}y_{a}\Big)^{N}\Big(\frac{q}{t}\Big)^{\frac{g-1}{2}N}\,\\\nonumber
L_{\vec{\nu}}(y)&=&\prod_{(i,j)\in \nu_{\alpha}}
[h_{\alpha\alpha}(i,j)]_{y^{-1}}
\prod_{\alpha<\beta}\Big(\prod_{(i,j)\in \nu_{\alpha}}[h_{\alpha\beta}(i,j)]_{y^{-1}Q_{\alpha\beta}}
\prod_{(i,j)\in \nu_{\beta}}[-\widetilde{h}_{\beta\alpha}(i,j)]_{y^{-1}Q_{\alpha\beta}}\Big)\\\nonumber
R_{\vec{\nu}}(y)&=& \prod_{\alpha=1}^{N}\prod_{(i,j)\in \nu_{\alpha}}
[\widetilde{h}_{\alpha\alpha}(i,j)]_{y}
\prod_{\alpha<\beta}\Big(\prod_{(i,j)\in \nu_{\alpha}}[\widetilde{h}_{\alpha\beta}(i,j)]_{y\,Q_{\alpha\beta}}
\prod_{(i,j)\in \nu_{\beta}}[-h_{\beta\alpha}(i,j)]_{yQ_{\alpha\beta}}\Big)\,.
\eea
$L_{\vec{\nu}}$ and $R_{\vec{\nu}}$ are defined in terms of relative hook lengths and quantum numbers which are defined as
\bea\nonumber
h_{\alpha\,\beta}(i,j)&=&(\nu_{\alpha,i}-j+1)-\tfrac{\epsilon_{2}}{\epsilon_{1}}(\nu^{t}_{\beta,j}-i)\,,
\widetilde{h}_{\alpha\,\beta}(i,j)=h_{\alpha\beta}(i,j)-1-\frac{\epsilon_{2}}{\epsilon_{1}}\\\nonumber
[x]&=&q^{x/2}-q^{-x/2}\,,\,\,\,\,[x]_{Q}=Q^{\frac{1}{2}}q^{x/2}-Q^{-\frac{1}{2}}q^{-x/2}\,.
\eea

In the limit $m_{a}\mapsto 0$ we get
\bea\label{PFrefined}
Z^{g}_{N}({\bf 0},\epsilon_1,\epsilon_2)&=&\sum_{\vec{\nu}}\widetilde{Q}^{|\vec{\nu}|}\,f_{\vec{\nu}}^{p+g-1}(L_{\vec{\nu}}(1)\,R_{\vec{\nu}}(1))^{g-1}\,,
\eea
and since for $\epsilon_{2}=-\epsilon_{1}$ we have $R_{\vec{\nu}}=L_{\vec{\nu}}$ therefore the partition function becomes,
\bea\label{PFUR}
Z^{g}_{N}({\bf 0},\epsilon_1,-\epsilon_1)&=&\sum_{\vec{\nu}}\widetilde{Q}^{|\vec{\nu}|}\,f_{\vec{\nu}}^{p+g-1}\,L_{\vec{\nu}}^{2g-2}\,.
\eea
In the section 4 and section 5 we will see that $Z^{g}_{N}({\bf 0},\epsilon_1,\epsilon_2)$ is precisely given by a topological field theory construction with TFT amplitudes being given by certain open topological string amplitudes.

\section{TFT Amplitudes for Local Curve: A Review}
In this section we will review, following \cite{BP,Aganagic:2004js}, the case of the local curve i.e., a Calabi-Yau threefold which is a rank two bundle over a curve of genus $g$ which corresponds to $N=1$. We denote by $X_{g}$ the Calabi-Yau threefold which is the total space of $L_{1}\oplus L_{2}\mapsto \Sigma_{g}$. $L_{1}$ and $L_{2}$ are line bundles on $\Sigma_{g}$, the genus $g$ Riemann surface, of degree $d_{1}$ and $d_{2}$ respectively. $X$ is a Calabi-Yau threefold if
\bea\nn
d_{1}+d_{2}=2g-2\,.
\eea
A well known $g=0$ case of this is the ${\cal O}(-1)\oplus {\cal O}(-1)\mapsto \mathbb{P}^{1}$, the resolved conifold. The  general $g=1$ case ${\cal O}(p)\oplus {\cal O}(-p)\mapsto T^{2}$ was also discussed in \cite{Vafa:2004qa}.

The Riemann surface $\Sigma_{g}$ can be obtained by gluing certain number of pair of pants, cylinders (which we will call the propagator) and the caps. When the Riemann surface is split into these basic building blocks the Calabi-Yau threefold in which the Riemann surface is embedded also splits into Calabi-Yau threefolds with boundaries. By placing Lagrangian branes on these boundaries of the Calabi-Yau threefold building blocks the amplitudes associated with each of these basic building blocks can be obtained using the open topological strings. This was the point of view taken in \cite{Aganagic:2004js}.  The amplitudes associated with the building blocks, coming from topological open strings amplitudes, can also be considered as amplitudes of a two dimensional QFT on a Riemann surface with boundaries. For the case of A-model amplitudes it was shown in \cite{Aganagic:2004js} that this 2D QFT is related to two dimensional q-deformed Yang-Mills theory.

If $\Sigma_{g,h}$ is a Riemann surface of genus $g$ with $h$ boundaries in the Calabi-Yau threefold $L_{1}\oplus L_{2}\mapsto \Sigma_{g,h}$ then the Calabi-Yau condition requires that \cite{BP}
\bea
d_{1}+d_{2}=2g-2+h\,.
\eea

In this section and later we will be using web diagrams to represent Calabi-Yau threefolds and Lagrangian branes. The relation between degeneration loci of toric Calabi-Yau threefolds and web diagrms was clarified and discussed in detail in \cite{Leung:1997tw} and also reviewed in \cite{Aganagic:2003db}. The role and description of Lagrangian branes in toric Calabi-Yau threefolds was discussed in \cite{Aganagic:2000gs}. We refer the reader to these papers for a discussion of toric Calabi-Yau threefolds and Lagrangina branes in them.

\subsection{Cap and Propagator Amplitude}

\begin{figure}[h]
  \centering
  \includegraphics[width=3.7in]{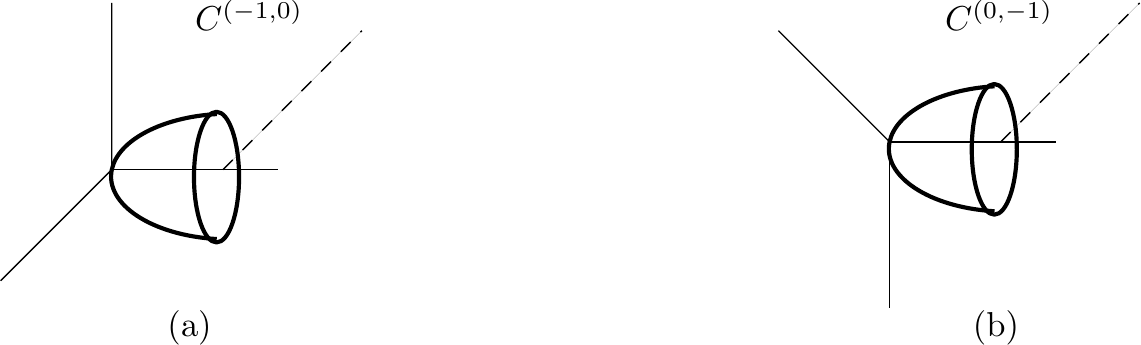}\\
  \caption{Two caps in $\mathbb{C}^3$ with different framing.}\label{caps}
\end{figure}

For the case of the cap we have $(g,h)=(0,1)$ therefore $d_{1}+d_{2}=-1$. We denote cap with degrees $(d_{1},d_{2})$ by $C^{(d_{1},d_{2})}$. Following the conventions of \cite{Aganagic:2004js} the cap amplitudes corresponding to $C^{(0,-1)}$ and $C^{(-1,0)}$ are given by
\bea
Z_{C}^{(0,-1)}(U)=\sum_{\lambda}C_{\lambda\emptyset\emptyset}\,\mbox{Tr}_{\lambda}U\,,
Z_{C}^{(-1,0)}(U)=\sum_{\lambda}C_{\lambda\emptyset\emptyset}\,q^{-\frac{\kappa(\lambda)}{2}}\,\mbox{Tr}_{\lambda}U\,,
\eea
where $C_{\lambda\mu\nu}$ is the topological vertex \cite{Aganagic:2003db} defined in Eq.(\ref{TVdefinition}) of Appendix B. In the above equations the holonomy on the boundary has been denoted by $U$. The cap amplitudes given above are open topological string amplitudes corresponding to the two toric geometries shown in \figref{caps}.

The two toric geometries shown in \figref{caps}(a) and \figref{caps}(b) differ from each other by an $SL(2,\mathbb{Z})$ transformation on the fibred torus which changes the framing of the Lagrangian brane. This change in framing appears as the factor $q^{-\frac{\kappa(\lambda)}{2}}$ in the amplitude associated with $C^{(-1,0)}$.

In the case that the Riemann surface is a cylinder so that $(g,h)=(0,2)$ we have $d_{1}+d_{2}=0$. We will denote the corresponding amplitude by $A^{(k,-k)}$. Since gluing a cylinder $A^{(0,0)}$ to either of the caps $C^{(-1,0)}$ or $C^{(0,-1)}$ should not change the caps, the amplitude associated with $A^{(0,0)}$ is
\bea
Z_{A}^{(0,0)}(U_{1},U_{2})=\sum_{\lambda}\mbox{Tr}_{\lambda}U_{1}\,\mbox{Tr}_{\lambda}U_{2}\,,
\eea
where $U_{1}$ and $U_{2}$ are the holonomies on the two boundaries of $A^{(0,0)}$. $A^{(-k,k)}$ can be obtained from $A^{(0,0)}$ by an $SL(2,\mathbb{Z})$ on the torus fibration of the Calabi-Yau threefold. This $SL(2,\mathbb{Z})$ transformation changes the framing of the Lagrangian branes and gives an addition factor related to the framing in the amplitude for $A^{(-k,k)}$ as compared with the amplitude $A^{(0,0)}$,
\bea\label{propagator}
Z_{A}^{(-k,k)}(U_{1},U_{2})=\sum_{\lambda}q^{-k\frac{\kappa(\lambda)}{2}}\,\mbox{Tr}_{\lambda}U_{1}\,\mbox{Tr}_{\lambda}U_{2}\,.
\eea
As discussed in \cite{Aganagic:2004js} we can insert an additional factor if the propagator (the cylinder) is has length $t$,
\bea\nonumber
Z_{A}^{(-k,k)}(U_{1},U_{2})=\sum_{\lambda}e^{-t|\lambda|}\,q^{-k\frac{\kappa(\lambda)}{2}}\,\mbox{Tr}_{\lambda}U_{1}\,\mbox{Tr}_{\lambda}U_{2}\,.
\eea
The cap $C^{(-k,k-1)}$ can be obtained by gluing $C^{(0,-1)}$ and $A^{(-k,k)}$ and has corresponding amplitude
\bea\label{cap}
Z_{C}^{(-k,k-1)}=\sum_{\lambda}C_{\lambda\emptyset\emptyset}\,q^{-k\frac{\kappa(\lambda)}{2}}\,\mbox{Tr}_{\lambda}U\,.
\eea
\vskip0.2cm
\subsubsection{${\cal O}(n-2)\oplus{\cal O}(-n)\mapsto \mathbb{P}^{1}$} Since the degrees of the bundles add under gluing of the base curves, we can construct the ${\cal O}(n-2)\oplus{\cal O}(-n)\mapsto \mathbb{P}^{1}$ by gluing $C^{(-m,m-1)}, A^{(-k,k)}$ and $C^{(-p,p-1)}$ such that $m+k+p=2-n$. The corresponding partition function is indeed the topological string partition function of ${\cal O}(n-2)\oplus {\cal O}(-n)\mapsto \mathbb{P}^{1}$ obtained from the topological vertex:
\bea\nn
Z^{(n-2,-n)}_{\mathbb{P}^{1}}&=&\int dU\,dV Z_{C}^{(-m,m-1)}(U)\,Z_{A}^{(-k,k)}(U^{-1},V)\,Z_{C}^{(-p,p-1)}(V^{-1})\,,\\\nn
&=&\sum_{\lambda}e^{-t|\lambda|}\,C_{\lambda\emptyset\emptyset}\,C_{\lambda\emptyset\emptyset}q^{-(m+p+k)\frac{\kappa(\lambda)}{2}}=\sum_{\lambda}e^{-t|\lambda|}\,C_{\lambda\emptyset\emptyset}\,C_{\lambda^{t}\emptyset\emptyset}\,\,q^{(n-1)\frac{\kappa(\lambda)}{2}}\,.
\eea

\subsubsection{${\cal O}(-n)\oplus{\cal O}(n)\mapsto T^2$} This can be obtained from just the propagator glued to itself. The corresponding partition function was discussed in detail in \cite{Vafa:2004qa} and is given by
\bea
Z^{(-n,n)}_{T^{2}}&=&\int dU Z_{A}^{(-n,n)}(U^{-1},U)\,,\\\nn
&=&\sum_{\lambda}e^{-t|\lambda|}\,q^{-n\frac{\kappa(\lambda)}{2}}\,.
\eea

\subsection{Pair of pants amplitude}
\begin{figure}[h]
  \centering
  \includegraphics[width=3in]{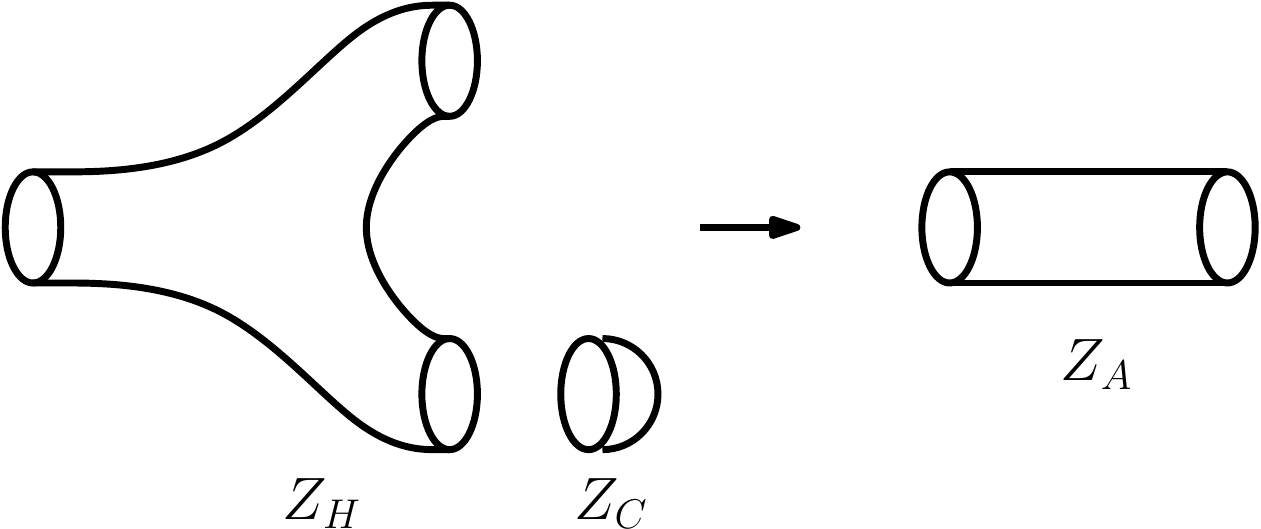}
  \caption{Gluing pair of pants and cap to obtain the propagator.}\label{pair}
\end{figure}

For the Riemann surface corresponding to the pair of pants $(g,h)=(0,3)$ and hence $d_{1}+d_{2}=1$. The pair of amplitude can be written as,
\bea\nn
Z^{(k,1-k)}_{H}(\vec{U})=\sum_{\lambda\,\mu\,\nu}Z^{(k,1-k)}_{\lambda\mu\nu}\mbox{Tr}_{\lambda}U_{1}\,\mbox{Tr}_{\mu}U_{2}\,\mbox{Tr}_{\nu}U_{3}\,.
\eea
The gluing of pair of pants with a cap gives the propagator as shown in \figref{pair} and this can be expressed as
\bea\nn
\int dU_{3}Z^{(k,1-k)}_{H}(\vec{U})Z_{C}^{(-m,m-1)}(U_{3})=Z^{(k-m,m-k)}_{A}(U_{1},U_{2})\,.
\eea

Eq.(\ref{propagator}) and Eq.(\ref{cap}) together with the above equation determines the pair of pants amplitude completely and it is given by,
\bea\label{ppp}
Z^{(k,1-k)}_{H}(\vec{U})=\sum_{\lambda}\,\frac{q^{k\frac{\kappa(\lambda)}{2}}}{C_{\lambda\emptyset\emptyset}}\,\mbox{Tr}_{\lambda}U_{1}\,\mbox{Tr}_{\lambda}U_{2}\,\mbox{Tr}_{\lambda}U_{3}\,.
\eea

\subsection{Genus creating operator}
The amplitude associated with the handle $(g,h)=(1,2)$ is also the matrix element of the genus creating operator. This amplitude can be obtained by gluing two pair of pants amplitudes as shown in \figref{GCO} below,
\bea
\int dU_{2}dU_{3}Z_{H}^{(k_{1},1-k_{1})}(U,U_{2},U_{3})Z_{H}^{(k_{2},1-k_{2})}(U_{2},U_{3},V)=G^{(k_{1}+k_{2},2-k_{1}-k_{2})}(U,V)
\eea
Using Eq.(\ref{ppp}) in the above equation we get,
\bea
G^{(k,2-k)}(U,V)&=&\sum_{\lambda}\,\frac{q^{k\frac{\kappa(\lambda)}{2}}}{C_{\lambda\emptyset\emptyset}^2}\,\mbox{Tr}_{\lambda}U\,\mbox{Tr}_{\lambda}V\,.
\eea

\begin{figure}[h]
  \centering
  \includegraphics[width=3.3in]{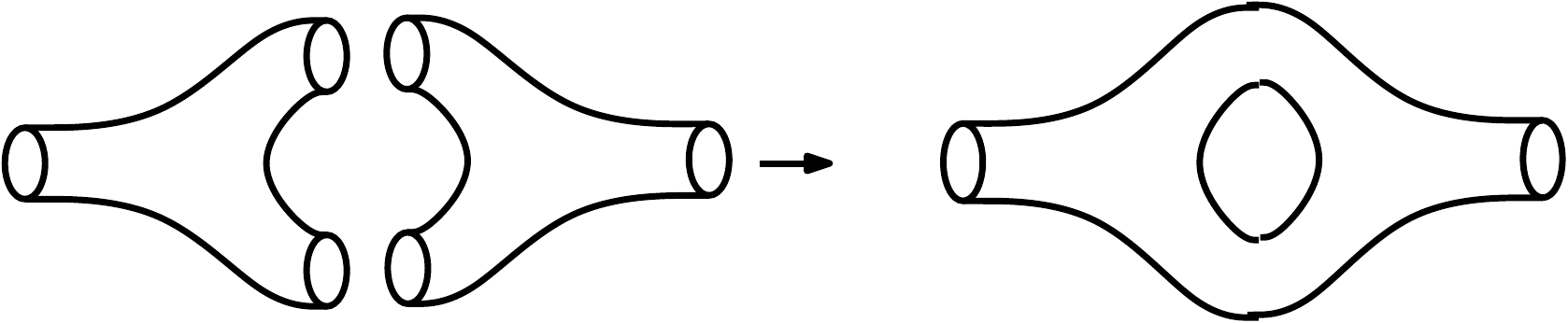}
  \caption{Gluing two pair of pants gives the amplitude associated with the genus creating operator.}\label{GCO}
\end{figure}

If we denote the genus creating operator by $\widehat{G}^{(k,2-k)}$ then its matrix element is given by,
\bea
\langle \lambda|\widehat{G}^{(k,2-k)}|\mu\rangle=\delta_{\lambda\mu}\frac{q^{k\frac{\kappa(\lambda)}{2}}}{C_{\lambda\emptyset\emptyset}^2}
\eea

Using the operator $\widehat{G}^{(k,2-k)}$ we can write down the partition function associated with a genus $g$ surface as a matrix element of product of genus creating operators with various framings. This essentially follows from the decomposition of the genus $g$ surface into $g$ handles and two caps,
\bea\nn
Z_{g}&=&\langle C^{a}|\underbrace{\widehat{G}^{(k_{1},2-k_{1})}\widehat{G}^{(k_{2},2-k_{2})}\cdots \widehat{G}^{(k_{g},2-k_{g})}}_{g \,\text{times}}|C^{b}\rangle\\\nn
&=&\sum_{\lambda}\frac{q^{f\frac{\kappa(\lambda)}{2}}}{C_{\lambda\emptyset\emptyset}^{2g-2}}
\eea
where $f=\sum_{i=1}^{g}k_{i}-a-b$ and $\langle \lambda|C^{a}\rangle=C_{\lambda\emptyset\emptyset}q^{-a\frac{\kappa(\lambda)}{2}}$.

Another way in which a genus $g$ Riemann surface can be obtained from handles is to decompose it into $g-1$ handles and one propagator (cylinder). In this case the partition function is the same as before since the underlying theory is topological and does not care how the Riemann surface is decomposed,
\bea\nn
Z_{g}&=&\mbox{Tr}\Big(\underbrace{\widehat{G}^{(k_{1},2-k_{1})}\widehat{G}^{(k_{2},2-k_{2})}\cdots \widehat{G}^{(k_{g-1},2-k_{g-1})}}_{g-1 \,\text{times}}\widehat{P}_{k}\Big)\\\label{PFRSUR}
&=&\sum_{\lambda}\frac{q^{f\frac{\kappa(\lambda)}{2}}}{C_{\lambda\emptyset\emptyset}^{2g-2}}
\eea
where $f=\sum_{i=1}^{g-1}k_{i}-k$ and $\widehat{P}_{k}$ is the operator corresponding to the framed propagator with matrix element $\langle \lambda|\widehat{P}_{k}|\mu\rangle =\delta_{\lambda\mu}q^{-k\frac{\kappa(\lambda)}{2}}$.

The partition function given by Eq.(\ref{PFUR}) for $N=1$ becomes,
\bea\nn
Z^{g}_{N=1}({\bf 0},\epsilon_{1},-\epsilon_{1})=\sum_{\nu}(\widetilde{Q}e_{1}^{-p-g+1})^{|\nu|}\,\frac{q^{-p\frac{\kappa(\nu)}{2}}}{C_{\emptyset\emptyset\nu}^{2g-2}}\,.
\eea
Comparing the above with Eq.(\ref{PFRSUR}) we see that two partition functions are the same for $\widetilde{Q}e_{1}^{-p-g+1}=1$ and $f=-p$.

\section{TFT Amplitudes for $A_{N-1}$ fibration: The Unrefined Case}
In the last section we reviewed the case of local curve and saw that the open string amplitudes can be thought of as TFT amplitudes associated with various pieces of a Riemann surface. The partition function associated with the genus $g$ curve obtained from these TFT amplitudes is equal to the supersymmetric partition function of the $U(1)$ gauge theory with $g$ massless adjoint hypermultiplets. In this section we generalize this result to the case of $U(N)$ gauge theory with $g$ massless adjoint hypermultiplets and show that the partition function of this theory can be obtained from TFT like amplitudes by gluing cap, propagator and pair of pants to form a genus $g$ Riemann surface.

The TFT like amplitudes associated with cap, propagator and pair of pants in the $U(N)$ case as well  are given by open string amplitudes on a Calabi-Yau threefold which is a fibration of resolved $A_{N-1}$ singularity over the cap, propagator and the pair of pant respectively. We will denote this Calabi-Yau threefold by $X^{N}_{C}, X^{N}_{A}$ and $X^{N}_{H}$ respectively. The three amplitudes can then be used to determine partition function associated with any Riemann surface with or without boundary. The Calabi-Yau threefolds $X^{N}_{C}, X^{N}_{A}$ and $X^{N}_{H}$ have boundaries and therefore one can associate open topological string amplitudes with these pieces.

In this section we will consider the unrefined case so that $\epsilon_{2}=-\epsilon_{1}$. The refined case and its amplitudes will be discussed in section V.

\subsection{The Cap}

\begin{figure}[h]
  \centering
  \includegraphics[width=4in]{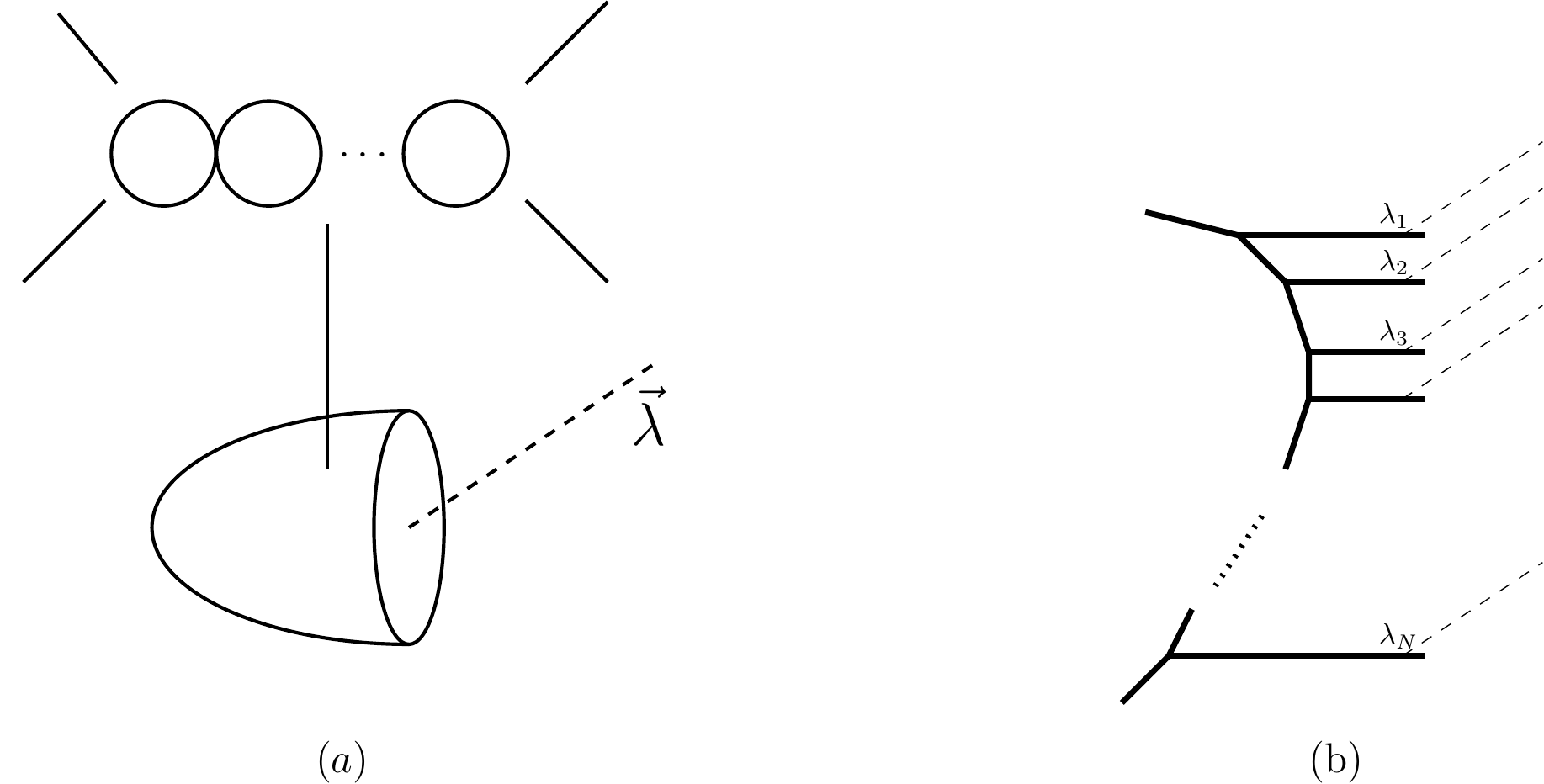}\\
  \caption{ (a) $\mbox{Cap}\times A_{N-1}$ with a stack of Lagrangian brane on the boundary of the cap. (b) The toric geometry of $\mbox{Cap}\times A_{N-1}$ with Lagrangian branes giving the cap amplitude.}\label{toric1}
\end{figure}

The Calabi-Yau threefold corresponding to the cap is $A_{N-1}\times \mathbb{C}$ with boundary of Lagrangian branes on $\mathbb{C}$. We place $N$ stacks of Lagrangian branes with topology $S^{1}\times \mathbb{R}^2$ in this geometry such that the $S^{1}$ of the Lagrangian branes wraps the topologically trivial $S^{1}$ in $\mathbb{C}$ and $\mathbb{R}^{2}$ extends in the resolved $A_{N-1}$ direction as shown in \figref{toric1}.

\begin{figure}[h]
  \centering
  \includegraphics[width=3.5in]{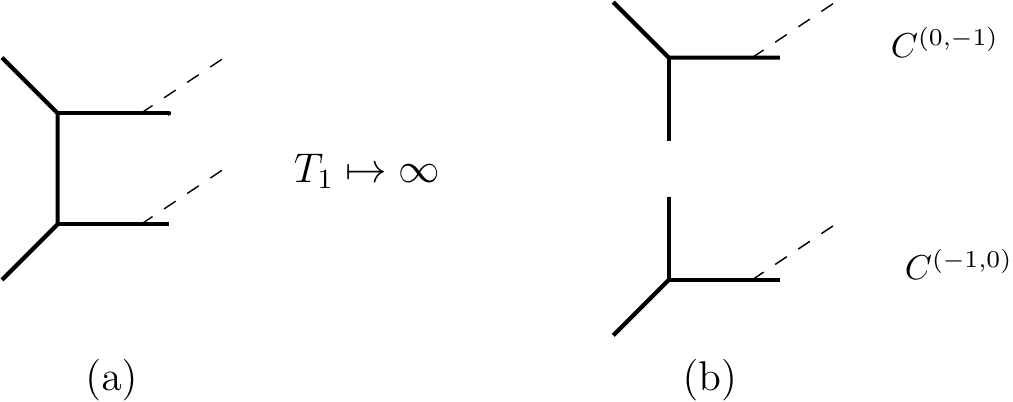}\\
  \caption{ (a) The geometry corresponding to cap amplitude for $N=2$. (b) In the limit $T\mapsto \infty$ this geomery splits in to two patches each a $\mathbb{C}^3$ with two Lagrangian branes on them with different relative framing.}\label{figgg}
\end{figure}

The caps in the geometry are labelled by $N$ pairs of numbers $(k_{a,1}, k_{a,2})$ such that $k_{a,1}+k_{a,2}=-1$. However, these numbers are not independent of each other, since the various $\mathbb{P}^1$'s in the fiber are glued together, and are determined by a single $k$ such that
\bea\label{framing1}
\underline{\bf k}:&=&(k_{a,1},k_{a,2})\\\nn
&=&(-k+1,k-1)+(a-2,-a+1)=(-k+a-1,k-a)\,,\,\,\,a=1,\cdots,N\,.
\eea
In the above equation the factor $(-k+1,k-1)$ comes from choosing a reference framing for the first curve and the fcator $(a-2,-a+1)$ is the intrinsic framing of the curve in our convention. To see this lets us consider the case of $N=2$ in which the $A_{1}$ geometry is shown in \figref{figgg}(a). If we take the size of the fiber $\mathbb{P}^1$ to infinity then the geometry decomposes into two patches each one a $\mathbb{C}^3$. From \figref{figgg}(b) we see that these two patches along with Lagrangian branes on them corresponds to the two caps shown in \figref{caps} with framing given by $(a-2,-a+1)$ for $a=1,2$ \cite{Aganagic:2003db}.

The corresponding cap amplitude can be calculated using the topological vertex associating framing determined by Eq.(\ref{framing1}) with each horizontal line and is given by,
\bea
Z_{C^{k}}({\bf U})=\sum_{\vec{\lambda}}Z^{k}_{\vec{\lambda}}\mbox{Tr}_{\vec{\lambda}}{\bf U}\,.
\eea
We have denoted the set of $N$ partitions $\{\lambda_1,\lambda_2,\cdots,\lambda_N\}$ by $\vec{\lambda}$, $\mbox{Tr}_{\vec{\lambda}}{\bf U}=\prod_{a=1}^{N}\mbox{Tr}_{\lambda_{a}}U_{a}$ and
\bea\label{cap0}
Z^{k}_{\vec{\lambda}}&=&\sum_{\vec{\mu}}\Big(\prod_{i=1}^{N-1}Q_{i}^{|\mu_{i}|}\Big)\,C_{\emptyset\mu_{1}\lambda_{1}}C_{\mu_{1}^{t}\mu_{2}\lambda_{2}}\cdots C_{\mu_{N-1}\emptyset\lambda_{N}}\,,
\eea
where $T_{i}=-\mbox{ln}(Q_{i})$ are the complexified K\"ahler parameters associated with the fiber $\mathbb{P}^{1}$'s. Using the expression for the refined topological vertex given by Eq.(\ref{TVdefinition}) and the Schur function identities given in Eq.(\ref{ID2}) we can write the open string amplitude in Eq.(\ref{cap0}) as a product,
\bea\label{cap1}
Z^{k}_{\vec{\lambda}}&=&G_{\vec{\lambda}}\frac{\prod_{a=1}^{N}\Big(q^{(-k+a-1)\frac{\kappa(\lambda_a)}{2}}\,C_{\lambda_{a}\emptyset\emptyset}\Big)}
{\prod_{1\leq a<b\leq N}N_{\lambda_{a}\lambda_{b}}(Q_{ab},\epsilon)}\,,
\eea
where $Q_{ab}=Q_{a}Q_{a+1}\cdots Q_{b-1}$ and we have defined an overall factor given by,
\bea
G_{\vec{\lambda}}=\Big(\prod_{\alpha<\beta}\,Q_{\alpha\beta}^{\frac{|\lambda_{\alpha}|+|\lambda_{\beta}|}{2}}\Big)
\Big(\prod_{\alpha=1}^{N}\,e_{\alpha}^{-|\lambda_{\alpha}|}\Big)^{k-\frac{N}{2}}\,,
\eea
which can be absorbed by redefining the holonomies but we will not do so for later convenience. 

Using the following identities,
\bea
C_{\lambda_{a}\emptyset\emptyset}&=&q^{\frac{\kappa(\lambda_a)}{4}}\,\prod_{(i,j)\in \lambda_{a}}\,[h(i,j)]^{-1}\\\nn
\prod_{1\leq a<b\leq N}N_{\lambda_{a}\lambda_{b}}(Q_{ab},\epsilon)&=&q^{\sum_{a=1}^{N}(a-\frac{N+1}{2})\frac{\kappa(\lambda_a)}{2}}]\,\prod_{1\leq a<b\leq N}\,Q_{ab}^{\frac{|\lambda_{a}|+|\lambda_b|}{2}}\,
\prod_{(i,j)\in \lambda_{a}}[h_{\lambda_{a}\lambda_{b}}]_{Q_{ba}}\,\prod_{(i,j)\in \lambda_{b}}[-h_{\lambda_{b}\lambda_{a}}]_{Q_{ba}}
\eea
we can write Eq.(\ref{cap1}) as
\bea
Z^{\underline{\bf k}}_{\vec{\lambda}}=\frac{f_{\vec{\lambda}}^{k-\frac{N}{2}}}{L_{\vec{\lambda}}(1)}
\eea
with
\bea
f_{\vec{\lambda}}:=\Big(\prod_{\alpha=1}^{N}\,e_{\alpha}^{-|\lambda_{\alpha}|}\,q^{-\frac{\kappa(\lambda_{a})}{2}}\Big)\,.
\eea
Thus the amplitude associated with the cap with framing $k$ is,
\bea
Z_{C^{k}}({\bf U})=\sum_{\vec{\lambda}}\frac{f_{\vec{\lambda}}^{k-\frac{N}{2}}}{L_{\vec{\lambda}}(1)}\,\mbox{Tr}_{\vec{\lambda}}{\bf U}\,,
\eea
Using the free fermionic Fock space ${\cal H}$ spanned by states labelled by partitions $|\vec{\lambda}\rangle$ \cite{Nekrasov:2003rj} we can write the above as state in ${\cal H}$,
\bea\label{capstate}
\boxed{|C^{(k)}\rangle=\sum_{\vec{\lambda}}\frac{f_{\vec{\lambda}}^{k-\frac{N}{2}}}{L_{\vec{\lambda}}(1)}\,|\vec{\lambda}\rangle\,.}
\eea
This state-operator notation for the amplitudes will be useful in the next section when we discuss the refined amplitudes for which $\epsilon_1+\epsilon_2\neq 0$.

\subsection{The Propagator}

The geometry corresponding to the propagator is $X^{N}_{A}:=A\times A_{N-1}$ where $A$ is a cylinder. If ${\bf U}=\{U_{1},\cdots U_{N}\}$ and ${\bf V}=\{V_{1},\cdots , V_{N}\}$ are the holonomies at the boundaries of $A$ then the amplitude associated with $A$ is given by,
\bea\nonumber
Z_{A}({\bf U},{\bf V})=\sum_{\vec{\lambda}\,\vec{\mu}}F_{\vec{\lambda}\vec{\mu}}(Q_{a})\,\mbox{Tr}_{\vec{\lambda}}{\bf U}\mbox{Tr}_{\vec{\mu}}{\bf V}\,,
\eea
where, as mentioned before, $T_{a}=-\mbox{ln}(Q_{a})$ are the complexified K\"ahler parameters associated with the fiber $\mathbb{P}^{1}$'s.
In the limit $Q_{a}\mapsto 0$ the Calabi-Yau geometry $X^{N}_{A}$ splits into $N$ copies of the $X^{1}_{A}$ discussed in the previous section with the corresponding amplitude becoming the identity. Therefore
\bea\label{p1}
F_{\vec{\lambda}\vec{\mu}}=\delta_{\vec{\lambda}\vec{\mu}}+{\cal O}(Q_{a})
\eea
Since two propagators can glued to obtain a new propagator therefore,
\bea\nonumber
\int d{\bf U}\,Z_{A}({\bf U},{\bf V})Z_{A}({\bf U}^{-1},{\bf W})=Z_{A}({\bf U},{\bf W})\,,
\eea
which implies,
\bea
F_{\vec{\lambda}\vec{\nu}}=\sum_{\vec{\mu}}F_{\vec{\lambda}\vec{\mu}}F_{\vec{\mu}\vec{\nu}}\,.
\eea
The above equation together with Eq.(\ref{p1}) gives $F_{\vec{\lambda}\vec{\mu}}=\delta_{\vec{\lambda}\vec{\mu}}$ and thus,
\bea
Z_{A}({\bf U},{\bf V})=\sum_{\vec{\lambda}}\,\mbox{Tr}_{\vec{\lambda}}{\bf U}\,\mbox{Tr}_{\vec{\lambda}}{\bf V}\,.
\eea

The propagator with framing can be obtained by introducing framing factor for each of the $N$ stack of branes on the boundary of the cylinder, which are separated in the fiber direction. The framed propagator is given by
\bea
Z_{A^{k}}({\bf U},{\bf V})=\sum_{\vec{\lambda}}f_{\vec{\lambda}}^{k}\,\mbox{Tr}_{\vec{\lambda}}{\bf U}\,\mbox{Tr}_{\vec{\lambda}}{\bf V}\,.
\eea
If we introduce a length for the propagator we can include that by scaling the holonomy matrix as discussed in \cite{Aganagic:2004js} to obtain,
\bea\label{modifiedpropagator}
\widehat{Z}_{A^{k}}({\bf U},{\bf V})=\sum_{\vec{\lambda}}f_{\vec{\lambda}}^{k}\,e^{-\vec{t}\cdot |\vec{\lambda}|}\,\,\mbox{Tr}_{\vec{\lambda}}{\bf U}\,\mbox{Tr}_{\vec{\lambda}}{\bf V}\,.
\eea
The operator acting on the Hilbert space ${\cal H}$ corresponding to the propagator is then given by
\bea\label{propagatoroperator}
\boxed{\widehat{A}^{(k)}=\sum_{\vec{\lambda}}f_{\vec{\lambda}}^{k}\,| \vec{\lambda}\rangle\langle \vec{\lambda}|\,.}
\eea

\subsection{The Pair of Pants}

The pair of pants amplitude can be determined from the cap and the propagator as shown in \figref{pair}. If we denote by $Z_{H^{(-k)}}({\bf U})=\sum_{\vec{\lambda}\vec{\mu}\vec{\nu}}Z^{(-k)}_{\vec{\lambda}\vec{\mu}\vec{\nu}}\mbox{Tr}_{\vec{\lambda}}\mbox{Tr}_{\vec{\mu}}\mbox{Tr}_{\vec{\nu}}$ the pair of pants amplitude then gluing the cap on the boundary with holonomy $U_{3}$ gives,
\bea
\int dU_{3}\,Z_{H^{(-k)}}(U_{1},U_{2},U_{3})\,Z_{C^{k}}(U_{3})=Z_{A^{0}}(U_{1},U_{2})\,.
\eea
Which implies,
\bea
\sum_{\vec{\nu}}Z^{-k}_{\vec{\lambda}\vec{\mu}\vec{\nu}}\,Z^{k}_{\vec{\nu}}=\delta_{\vec{\lambda}\vec{\mu}}\,,
\eea
Similarly gluing the cap onto the other two boundaries of the pair of pants give two more conditions which together with the above imply that the pair of pants is given by,
\bea
Z_{H^{(-k)}}({\bf U})=\sum_{\vec{\lambda}}(Z^{k}_{\vec{\lambda}})^{-1}\mbox{Tr}_{\vec{\lambda}}{\bf U}_{1}\mbox{Tr}_{\vec{\lambda}}{\bf U}_{2}\mbox{Tr}_{\vec{\lambda}}{\bf U}_{3}\,.
\eea
As a state in the tensor product ${\cal H}^{*}\otimes {\cal H}\otimes {\cal H}$ it is given by
\bea\label{pairofpantsstate}
\boxed{\widehat{H}^{(-k)}=\sum_{\vec{\lambda}}\,(Z^{k}_{\vec{\lambda}})^{-1}\,(\langle \vec{\lambda}|)\,\otimes\,|\vec{\lambda}\rangle\,|\vec{\lambda}\rangle\,.}
\eea

\begin{figure}[h]
  \centering
  \includegraphics[width=4in]{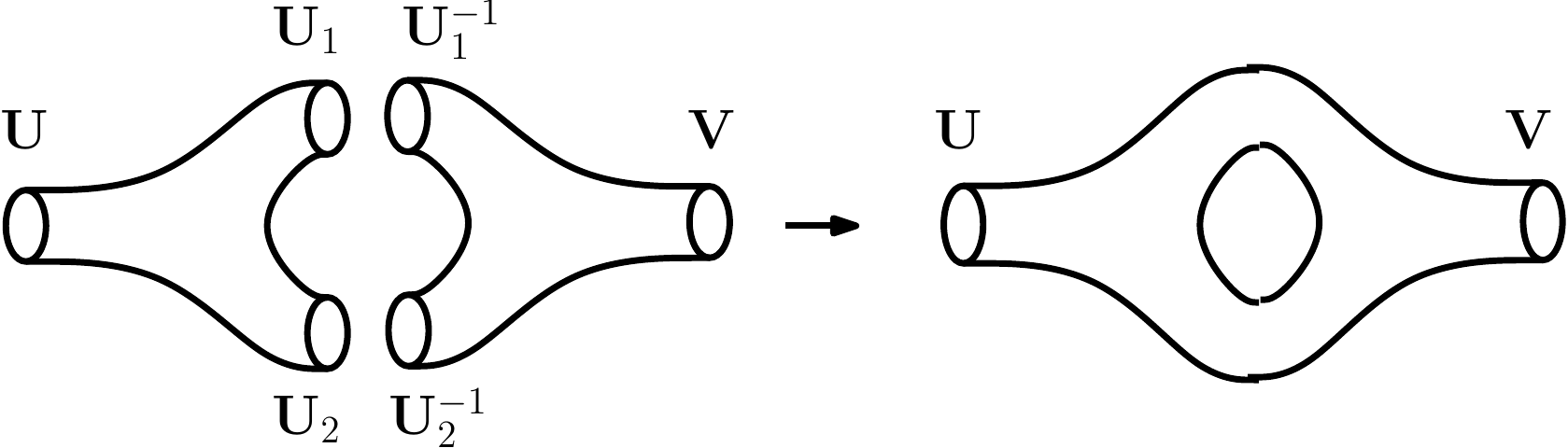}\\
  \caption{Amplitude associated with the surface of genus one with two boundaries can be obatined by gluing two pair of pants amplitudes. }\label{genus2}
\end{figure}

By gluing two pair of pants amplitude we can obtain the matrix element of the genus creating operator as shown in \figref{genus2} and also discussed in section 3 for $N=1$ case. 

The amplitude obtained by gluing two pair of pants according to \figref{genus2} is given by,
\bea
Z_{G}({\bf U},{\bf V})=\sum_{\vec{\lambda}}(Z^{k}_{\vec{\lambda}})^{-2}
\mbox{Tr}_{\vec{\lambda}}{\bf U}\,\mbox{Tr}_{\vec{\lambda}}{\bf V}\,.
\eea
If we denote by $\widehat{G}^{(k)}$ the genus creating operator with framing $k$ then its matrix element is given by,
\bea\label{genusoperator}
\boxed{\widehat{G}^{(k)}=\sum_{\vec{\lambda}}\,f^{k}_{\vec{\lambda}}(Z^{0}_{\vec{\lambda}})^{-2}\,|\vec{\lambda}\rangle\langle\vec{\lambda}|\,=\sum_{\vec{\lambda}}\,f^{k+N}_{\vec{\lambda}}\,(L_{\vec{\lambda}}(1))^2\,|\vec{\lambda}\rangle\langle\vec{\lambda}|\,.}
\eea

\subsubsection{Example 1: $A_{N-1}$ fibred over $\mathbb{P}^1$}

\begin{figure}[h]
 \centering
\includegraphics[width=2.5in]{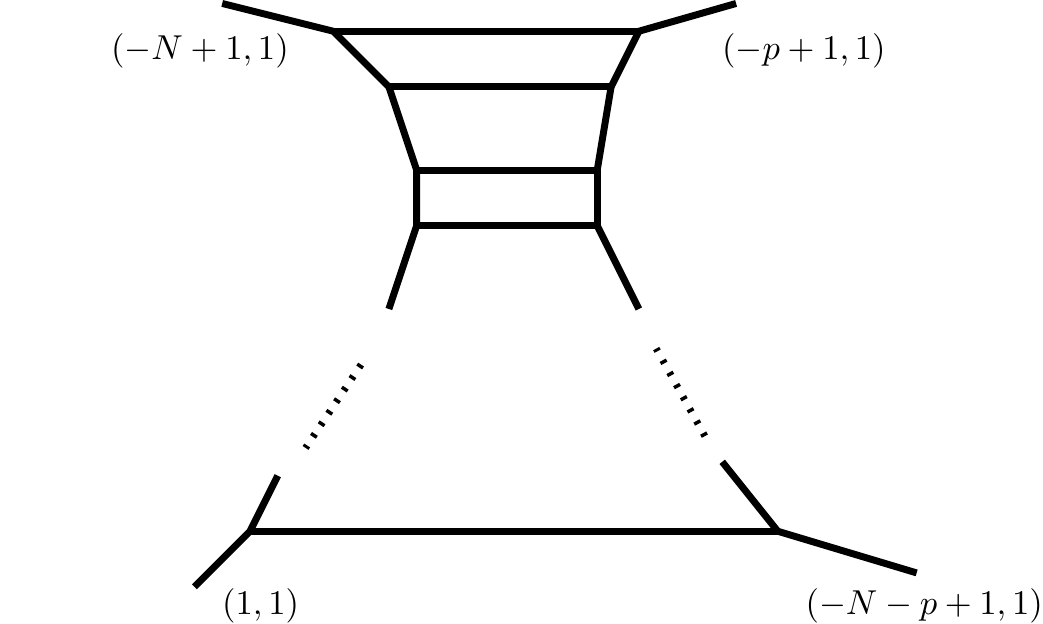}\\
 \caption{The web diagram of Calabi-Yau threefold which are resolved $A_{N-1}$ fibration over $\mathbb{P}^1$.}\label{toric2}
\end{figure}

We can obtain the partition function associated with $\mathbb{P}^1$ by gluing two disks together using a propagator:
\bea
Z_{\mathbb{P}^1}=\langle C^{(k_{1})}|\widehat{A}^{(k_{2})}\,|C^{(k_{3})}\rangle\,.
\eea
Using Eq.(\ref{capstate}) and Eq.(\ref{propagatoroperator}) we get
\bea\nn
&&Z_{\mathbb{P}^1}=\sum_{\vec{\lambda}}G_{\vec{\lambda}}^{2}\,f^{h}_{\vec{\lambda}}\,\frac{\prod_{a=1}^{N}q^{(2a-2-k-r)\frac{\kappa(\lambda_{a})}{2}}\,C_{\lambda_{a}\emptyset\emptyset}C_{\lambda_{a}\emptyset\emptyset}}
{\prod_{1\leq a<b\leq N}N_{\lambda_{a}\lambda_{b}}(Q_{ab})^2}\\\nn
&=&\sum_{\vec{\lambda}}\frac{f^{p-1}_{\vec{\lambda}}}{L_{\vec{\lambda}}(1)^2}
\eea
This is precisely the partition function given in Eq.(\ref{PFUR}) where $p-1=k_{1}+k_{2}+k_{3}-N$. If we introduce the modified propagator given in Eq.(\ref{modifiedpropagator}) then we get,
\bea\nn
Z_{\mathbb{P}^1}&=&\sum_{\vec{\lambda}}e^{-t|\vec{\lambda}|}G_{\vec{\lambda}}^{2}\,f^{h}_{\vec{\lambda}}\,\frac{\prod_{a=1}^{N}q^{(2a-2-k-r)\frac{\kappa(\lambda_{a})}{2}}\,C_{\lambda_{a}\emptyset\emptyset}C_{\lambda_{a}\emptyset\emptyset}}
{\prod_{1\leq a<b\leq N}N_{\lambda_{a}\lambda_{b}}(Q_{ab})^2}\\\nn
&=&\sum_{\vec{\lambda}}e^{-t|\vec{\lambda}|}\,\frac{f^{p-1}_{\vec{\lambda}}}{L_{\vec{\lambda}}(1)^2}\,.
\eea
This is precisely the supersymmetric partition function of the pure $U(N)$ gauge theory with $-p$ being the Chern-Simons coefficient. This is also the topological string partition function of the a Calabi-Yau threefold which is a resolved $A_{N-1}$ fibration over $\mathbb{P}^1$. There are $N+1$ such fibration labelled by the Chern-Simons coefficient $-p=0,1,2,\cdots,N$. The web diagram of the corresponding Calabi-Yau threefolds is shown in \figref{toric2}.
\vskip 0.07cm

In the above diagram we can see that the compact part of the Calabi-Yau threefold is obtained by gluing $N-1$ Hirzebruch surfaces. The local geometry of the base $\mathbb{P}^1$ in the $a-1$-th  and $a$-th Hirzebruch surface is given by,
\bea
{\cal O}(-N-p+2a-2)&\oplus& {\cal O}(N+p-2a)\\\nn
&\downarrow&\\\nn
&\mathbb{P}^1&
\eea
with $a=1,2,\cdots N$.

\subsubsection{Example 2: $A_{N-1}$ fibred over $T^2$}
The $T^2$ partition function can be obtained in two different ways. It can be obtained by gluing the two boundaries of the propagator,
\bea
Z_{T^2}=\mbox{Tr}\,\widehat{A}^{(k)}=\sum_{\vec{\lambda}}\,f^{k}_{\vec{\lambda}}
\eea
and it can also be obtained by gluing the caps onto the $(g,h)=(1,2)$ Riemann surface whose amplitude is given by the genus creating operator,
\bea
Z_{T^2}&=&\langle C^{(k_1)}|\widehat{G}^{(k_2)}|C^{(k_3)}\rangle\,,\\\nn
&=&\sum_{\vec{\lambda}}\,f^{k_1+k_2+k_3}_{\vec{\lambda}}\,.
\eea

\subsubsection{Example 3: $A_{N-1}$ fibred over $\Sigma_{g}$}
The partition function associated with $\Sigma_g$, genus $g$ surface, can be obtained in many different ways by gluing the three building blocks: cap, propagator and the pair of pants. For example, by gluing $2g$ pair of pants and then gluing two caps onto the resulting surface we can obtain $\Sigma_g$ with corresponding partition function,
\bea
Z_{\Sigma_g}&=&\langle C^{(k)}|\widehat{G}^{(k_1)}\cdots \widehat{G}^{(k_g)}|C^{(\ell)}\rangle\,,\\\nn
&=&\sum_{\vec{\lambda}}f^{h}_{\vec{\lambda}}\,(L_{\vec{\lambda}}(1))^{2g-2}\,,
\eea
where $h=k+\ell+(k_1+\cdots+k_{g})+(g-1)N$. We can also obtain $\Sigma_g$ by gluing $2g-2$ pair of pants and then gluing the remaining two boundaries by a propagator,
\bea
Z_{\Sigma_g}&=&\mbox{Tr}\Big(\widehat{G}^{(k_1)}\cdots \widehat{G}^{(k_{g-1})}\,\widehat{A}^{(k)}\Big)\,,\\\nn
&=&\sum_{\vec{\lambda}}f^{s}_{\vec{\lambda}}\,(L_{\vec{\lambda}}(1))^{2g-2}\,,
\eea
where $s=k+(k_1+\cdots+k_{g-1})+(g-1)N$. The above partition function is the same as Eq.(\ref{PFUR}) for $\widetilde{Q}=1$ and $p=k+(k_1+\cdots k_{g-1})+(g-1)(N-1)$.

\section{TFT Amplitudes for $A_{N-1}$ fibration: The Refined Case}

In this section we will discuss the refined case when $\epsilon_{1}+\epsilon_{2}\neq 0$ i.e., $q\neq t$. Apart from amplitudes depending on both $q$ and $t$ this case will be different from the unrefined case in two other ways. The cap amplitude was obtained by slicing the base $\mathbb{P}^1$ into two caps and taking the open string amplitude associated with one of the caps as shown in \figref{toric1}. In the refined case both the left and the right cap give the same amplitude, however, in the refined case the two amplitudes are not the same. For this reason we will introduce the left and the right cap and the corresponding amplitudes. The other difference will be the holonomy factor. In the unrefined case we had the factor $\mbox{Tr}_{\lambda}U$ for each brane. The factor $\mbox{Tr}_{\lambda}U=s_{\lambda}({\bf x})$ where ${\bf x}$ is the set of eigenvalues of the holonomy matrix $U$. In the refined case this factor changes from the Schur polynomials of the eigenvalues, to the Macdonald polynomial of  eigenvalues \cite{Iqbal:2011kq, Aganagic:2012hs}. The Macdonald polynomials $P_{\lambda}$ and their dual, with respect to the Macdonald inner product, $Q_{\lambda}$ form the correct basis for the expansion of refined open topological string partition function as was shown in \cite{Iqbal:2011kq, Aganagic:2012hs}. To write down the TFT amplitudes we will consider a Fock space such that $P_{\vec{\lambda}}(U;q,t)$ are the wavefunctions corresponding to the states and $|\vec{\lambda}\rangle\in {\cal H}$ and $Q_{\vec{\lambda}}(U^{-1};q,t)$ are the wavefunctions for the dual states $\langle \vec{\lambda}|\,\in {\cal H}^{*}$ such that $\langle \vec{\lambda}|\vec{\mu}\rangle=\delta_{\vec{\lambda}\vec{\mu}}$. The various states and operators are then given by:
\bea
\mbox{Right Cap:}&&\,\,|C^{R,k}\rangle \in {\cal H}\\\nn
\mbox{Left Cap:}&&\,\,\langle C^{L,k}| \in {\cal H}^{*}\\\nn
\mbox{Propagator:}&&\,\,\widehat{A}\in {\cal H}\otimes {\cal H}^{*}\\\nn
\mbox{Left Pair of Pants:}&&\,\,\widehat{H}_{L}\in  {\cal H}\otimes {\cal H}^{*}\otimes {\cal H}^{*}\\\nn
\mbox{Right Pair of Pants:}&&\,\,\widehat{H}_{R}\in {\cal H}\otimes {\cal H}\otimes {\cal H}^{*}\\\nn
\mbox{Genus Creating Operator:}&&\,\,\widehat{G}\in {\cal H}\otimes {\cal H}^{*}
\eea

\subsection{The Cap Amplitudes}
\begin{figure}[h]
 \centering
\includegraphics[width=4.5in]{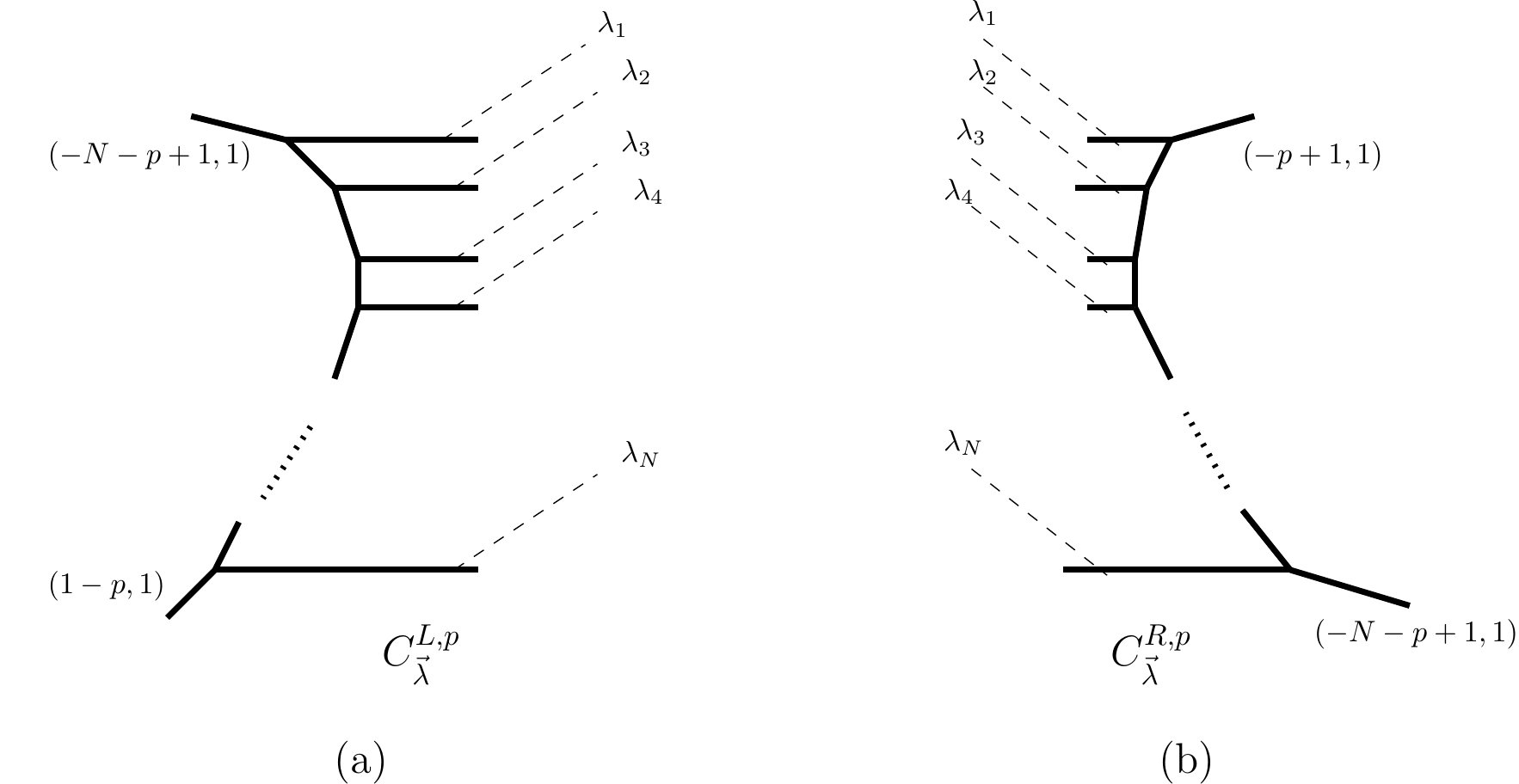}\\
 \caption{(a) The left cap and (b) the right cap.}\label{toric3}
\end{figure}

The cap amplitudes are obtained by considering the Calabi-Yau threefold which is $A_{N-1}$ space fibered over $\mathbb{P}^1$ and slicing the base $\mathbb{P}^1$ to two caps as shown in \figref{toric3}. The cap amplitudes $C^{R,k}_{\vec{\lambda}}$ and $C^{L,k}_{\vec{\lambda}}$ are the open topological string amplitudes associated with the left and the right portions if we cut the of the web diagram shown in \figref{toric2} in half by dividing the $N$ horizontal lines by a vertical line. 

The left and the right cap amplitudes can be calculated using the refined topological vertex \cite{Iqbal:2007ii} and are given by,
\bea\nonumber
C^{R,k}_{\vec{\lambda}}(t,q)=\sum_{\mu_{1},\cdots,\mu_{N}}\prod_{a=1}^{N}\Big[(-Q_{f_{a}})^{|\mu_{a}|}\,f_{\lambda_{a}}(t,q)^{k-a+1}f_{\mu_{a}}(t,q)\,C_{\mu_{a-1}\mu^{t}\lambda_{a}}(t,q)\Big]\,,
\eea
with $f_{\lambda}(t,q)=(\frac{t}{q})^{\frac{||\lambda^t||^2}{2}}\,q^{-\frac{\kappa(\lambda)}{2}}$ and $C_{\lambda\mu\nu}(t,q)$ is the refined topological vertex given in Eq.(\ref{TVdefinition}) in Appendix A. 
The left cap can similarly be calculated using \figref{toric3}(a). After some simplification they can be written as:
\bea\nonumber
C^{R,k}_{\vec{\lambda}}&=&G_{\vec{\lambda}}\frac{\prod_{a=1}^{N}f_{\lambda_{a}}(t,q)^{k-a+1}C_{\emptyset\emptyset\lambda_{a}}(t,q)}
{\prod_{1\leq a<b\leq N}\prod_{(i,j)\in \lambda_{a}}(1-Q_{ab}q^{\lambda_{a,i}-j}\,t^{\lambda^{t}_{b,j}-i+1})\prod_{(i,j)\in \lambda_{b}}(1-Q_{ab}t^{-\lambda^{t}_{a,j}+i}\,q^{-\lambda_{b,i}+j-1})}\\\label{x1}
&=&\frac{f_{\vec{\lambda}}^{k-\frac{N}{2}}}{R_{\vec{\lambda}}(1)}\,,
\eea
and
\bea\nonumber
C^{L,k}_{\vec{\lambda}}&=&G_{\vec{\lambda}}\frac{\prod_{a=1}^{N}f_{\lambda_{a}}(t,q)^{k-a}C_{\emptyset\emptyset\lambda^{t}_{a}}(q,t)}
{\prod_{1\leq a<b\leq N}\prod_{(i,j)\in \lambda_{a}}(1-Q_{ab}q^{\lambda_{a,i}-j+1}\,t^{\lambda^{t}_{b,j}-i})\prod_{(i,j)\in \lambda_{b}}(1-Q_{ab}t^{-\lambda^{t}_{a,j}+i-1}\,q^{-\lambda_{b,i}+j})}\,,\\\label{x2}
&=&\frac{f_{\vec{\lambda}}^{k-\frac{N}{2}}}{L_{\vec{\lambda}}(1)}\,.
\eea

We define the corresponding states as,
\bea\label{caprefined}
|C^{R,k}\rangle=\sum_{\vec{\lambda}}C_{\vec{\lambda}}^{R,k}\,|\vec{\lambda}\rangle\,,\,\,\,
\langle C^{L,k}|=\sum_{\vec{\lambda}}C^{L,k}_{\vec{\lambda}}\,\langle \vec{\lambda}|\,,
\eea

\subsection{The Propagator, the Pair of Pants and the Genus Creating Operator}
The argument similar to the one for the unrefined case shows that the propagator in the refined case is also diagonal. The framed propagator is then given by,
\bea\label{propagatorrefined}
\widehat{A}^{(k)}=\sum_{\vec{\lambda}}f^{k}_{\vec{\lambda}}\,|\vec{\lambda}\rangle\langle\vec{ \lambda}|\,.
\eea
Unlike the unrefined case their are two distinct pair of pants which we will denote with $H_{L}$ and $H_{R}$ such that
\bea\label{pairrefined}
H^{k}_{L}\,|C^{R,0}\rangle=\widehat{A}^{(k)}\,,\,\,\,\,\langle C^{L,0}|H^{k}_{R}=\widehat{A}^{(k)}\,.
\eea
\begin{figure}[h]
\centering
\includegraphics[width=4.25in]{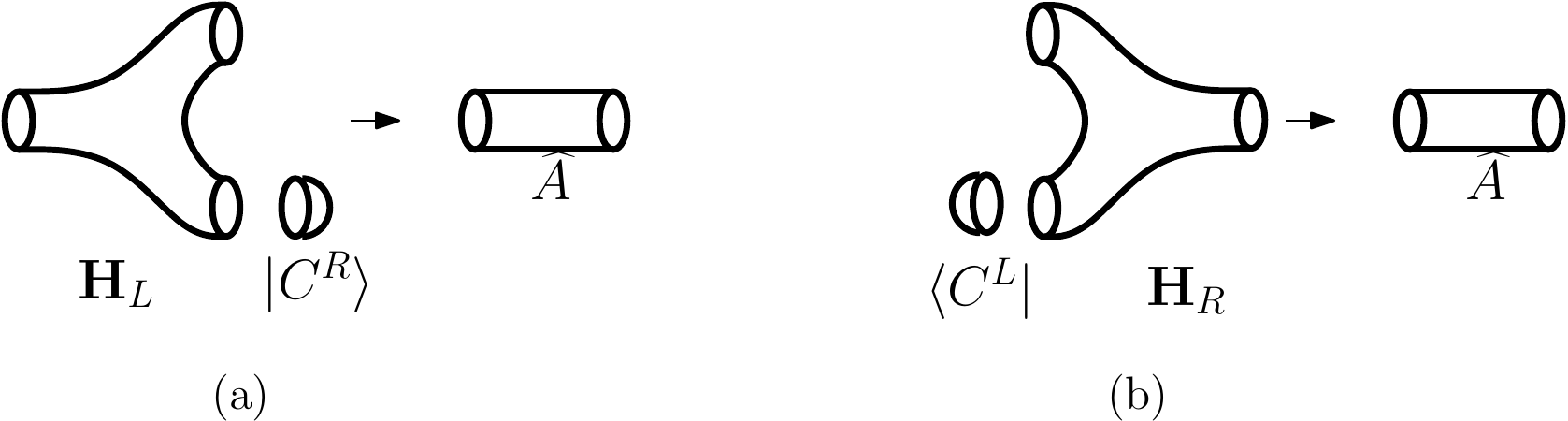}\\
 \caption{Two distinct pair of pants $H_{L}$ and $H_{R}$.}\label{pair2}
\end{figure}

Using Eq.(\ref{x1}), Eq(\ref{x2}) and Eq.(\ref{propagatorrefined}) together with Eq.(\ref{pairrefined}) gives,
\bea\nn
H^{k}_{L}=\sum_{\vec{\lambda}}\frac{f^{k}_{\vec{\lambda}}}{C^{R,0}}\,|\vec{\lambda}\rangle\,\langle \vec{\lambda}|\langle\vec{\lambda}|\,,\,\,H^{k}_{R}=\sum_{\vec{\lambda}}\frac{f^{k}_{\vec{\lambda}}}{C^{L,0}}\,|\vec{\lambda}\rangle\,| \vec{\lambda}\rangle\langle\vec{\lambda}|\,.
\eea

By gluing the left and the right pair of pants we can obtain the $(g,h)=(1,2)$ surface corresponding to the genus creating operator,
\bea
\widehat{G}^{(k)}&=&\sum_{\vec{\lambda}}\frac{f^{k}_{\vec{\lambda}}}{C^{R,0}C^{L,0}}\,|\vec{\lambda}\rangle\langle\vec{\lambda}|\,,\\\nn
&=&\sum_{\vec{\lambda}}f_{\vec{\lambda}}^{k+N}\,L_{\vec{\lambda}}(1)R_{\vec{\lambda}}(1)|\vec{\lambda}\rangle\langle\vec{\lambda}|\,.
\eea

\subsubsection{Example 1: $A_{N-1}$ fibration over $\mathbb{P}^{1}$}
To obtain the genus zero partition function we can glue the left and the right cap using a propagator,
\bea
Z_{\mathbb{P}^1}&=&\langle C^{L,k_{1}}|\widehat{A}^{k_{2}}|C^{R,k_{3}} \rangle\,,\\\nn
&=&\sum_{\vec{\lambda}}\frac{f_{\vec{\lambda}}^{k_{1}+k_{2}+k_{3}-N}}{L_{\vec{\lambda}}(1)R_{\vec{\lambda}}(1)}\,.
\eea
This precisely gives the partition function in Eq.(\ref{PFrefined}) for $\widetilde{Q}=1$ and $p-1=k_1+k_2+k_3-N$. The modified propagator can be used to introduce $\widetilde{Q}$ hence obtaining the refined topological string partition function of the Calabi-Yau threefold with web diagram given in \figref{toric2}.

\subsubsection{Example 2: $A_{N-1}$ fibration over genus $g$}
The refined partition function associated with $\Sigma_g$ is given by,
\bea\label{genusgPF1}
Z_{\Sigma_g}&=&\mbox{Tr}\Big(\widehat{G}^{(k_{1})}\widehat{G}^{(k_{2})}\cdots \widehat{G}^{(k_{g-1})}\,\widehat{A}^{k}\Big)\\\nn
&=&\sum_{\vec{\lambda}}f_{\vec{\lambda}}^{p+g-1}\,\Big(L_{\vec{\lambda}}(1)R_{\vec{\lambda}}(1)\Big)^{g-1}
\eea
where $p=k+\sum_{i=1}^{g-1}k_{i}+(g-1)(N-1)$. As discussed in last section $Z_{\Sigma_g}$ can also be obtained by gluing left/right pair of pants together with left/right caps,
\bea\label{genusgPF2}
Z_{\Sigma_g}&=&\langle C^{L,k}|\widehat{G}^{(k_{1})}\cdots \widehat{G}^{(k_{g})}|C^{R,\ell}\rangle\\\nn
&=&\sum_{\vec{\lambda}}f^{p+g-1}_{\vec{\lambda}}\Big(L_{\vec{\lambda}}(1)R_{\vec{\lambda}}(1)\Big)^{g-1}\,,
\eea
where $p=k+\ell+(g-1)(N-1)+\sum_{i=1}^{g}k_{i}$. Both Eq.(\ref{genusgPF1}) and Eq.(\ref{genusgPF2}) give the same result which is also equal to Eq.(\ref{PFrefined}) for $\widetilde{Q}=1$.

The refined $N=1$ case was also discussed in \cite{Aganagic:2012hs} where it was shown that in the refined case the two dimensional topological field theory is the two dimensional $(q,t)$-deformed Yang-Mills.

\subsection{Beyond Topological}
So far we have considered partition function with $m_{a}=0$ which was needed for the corresponding amplitudes to be understood as amplitudes of some underlying topological field theory. With $m_a=0$ we can glue the basic building blocks to form the Riemann surface in anyway but getting the same partition function.

For $m_a=0$ we saw in Eq.(\ref{genusgPF1}) that one representation of the partition function associated with $\Sigma_g$ is as a trace of product of genus creating operators. For generic $m_{a}$ we can also write down the partition function as a trace of product of genus creating operators which depends on $m_{a}$. Define the genus creating operator which depends on $m$ as follows,
\bea\nn
\widehat{G}^{(k)}(m)=\sum_{\vec{\lambda}}f_{\vec{\lambda}}^{k+N}\,L_{\vec{\lambda}}(e^{2\pi i m})\,R_{\vec{\lambda}}(e^{2\pi i m})|\vec{\lambda}\rangle\langle\vec{\lambda}|\,,
\eea
where $m=0$ gives the operator we discussed earlier. Also define the modified propagator which also depends on $m$ as follows,
\bea\nn
\widehat{A}^{(k)}(m)=\sum_{\vec{\lambda}}\,e^{-\vec{t}\cdot\vec{\lambda}}\,f^{k}_{\vec{\lambda}}\,\frac{L_{\vec{\lambda}}(e^{2\pi i m})\,R_{\vec{\lambda}}(e^{2\pi i m})}{L_{\vec{\lambda}}(1)\,R_{\vec{\lambda}}(1)}|\vec{\lambda}\rangle\langle\vec{\lambda}|\,.
\eea
The modified propagator is the refined open topological string amplitude of the geometry obtained by blowing up $N$ times the geometry shown in \figref{toric1} and assigning the same size $m$ to each blowup.

%

The partition function of the $U(N)$ gauge theory with $g$ adjoint hypermultiplets of mass $m_a$ given by Eq.(\ref{PF2}) can then be written as,
\bea\nn
Z_{g}&=&\mbox{Tr}\Big(\widehat{G}^{(k_{1})}(m_{1})\cdots G^{(k_{g-1})}(m_{g-1})\,\widehat{A}^{(k)}(m_{g})\Big)\\\nn
&=&\sum_{\vec{\lambda}}\,e^{-\vec{t}\cdot\vec{\lambda}}\,f^{p+g-1}_{\vec{\lambda}}
\frac{\prod_{a=1}^{g}L_{\vec{\lambda}}(e^{2\pi i m_a})\,R_{\vec{\lambda}}(e^{2\pi i m_a})}{L_{\vec{\lambda}}(1)\,R_{\vec{\lambda}}(1)}\,,
\eea
where $p=k+(g-1)(N-1)+\sum_{i=1}^{g-1}k_{i}$. The $g=1$ case was also discussed in detail in \cite{Iqbal:2008ra}.

\section{Conclusions}

In this paper we considered a decomposition of the resolved $A_{N-1}$ fibration over $\Sigma_{g}$, a Calabi-Yau threefold, given by the decomposition of $\Sigma_{g}$ in to cap, propagator and the pair of pants. Using open topological strings we associated with the cap, cylinder and pair of pants pieces of the Calabi-Yau threefold certain TFT-like amplitudes. We have shown that the partition function of the $U(N)$ gauge theory with $g$ massless adjoint hypermultiplets can be expressed in terms of these TFT-like amplitudes using the rules of the two dimensional topological field theory. This generalized the case of $N=1$ studied in \cite{BP,Aganagic:2004js}.

It will interesting to study the relation of the correlation function of chiral operators in the gauge theory with the correlation functions in the two dimensional topological field theory. For the $N=1$ case when the gauge group is $U(1)$ the corresponding topological field theory is known to be the q-deformed Yang-Mills theory \cite{Aganagic:2004js}. For $N>1$ the corresponding topological theory is q-deformed quiver Yang-Mills theory whose details and relation with black hole physics will be discussed elsewhere \cite{future}.

\section*{Acknowledgements}

We would like to thank Shehryar Sikander for participating in the initial stages of this project. We would also like to thank Duiliu-Emanuel Diaconescu, Soo-Jong Rey, Stefan Hohenegger, Can Kozcaz and Cumrun Vafa for useful discussions. A.I. and B.Q. were supported in part by the Higher Education Commission grant HEC-20-2518. AI was also supported by the Cheng fellowship from Center of Mathematical Sciences and Applications at Harvard University.

\section*{Appendix A}
In this section, we  collect some useful formulas which have been used in text. Throughout the text we have denoted by Greek letters $\lambda,\mu,\nu$ partitions of natural numbers and the notation $(i,j)\in \mu$ simply means the $(i,j)$ box in the Young diagram corresponding to the partition $\mu$. The index $i$ takes values in the set $\{1,2,\cdots,\ell(\mu)\}$ where $\ell(\mu)$ is the number of parts of the partition. For a fixed $i$ the index $j$ then takes values in the set $\{1,2,\cdots,\mu_{i}\}$ where $\mu_i$ is the $i$-th part of the partition. Given a partition $\mu$ we denote by $\mu^t$ the transpose of the partition obtained by reflection of the Young diagram of $\mu$ in the diagonal.

We used several functions defined on the partitions which we define below:
\bea
|\mu|&=&\sum_{i=1}^{\ell(\mu)}\mu_i\,,\,\,\,\,||\mu||^2=\sum_{i=1}^{\ell(\mu)}\mu^{2}_i\,\\\nn
\kappa(\mu)&=&||\mu||^2-||\mu^t||^2\,.
\eea

We interchangeably used $(\epsilon_1,\epsilon_2)$ and $(q,t)$ at several places in the text. The relation between them is given by:
\bea\nn
(q,t)=(e^{i\epsilon_1},e^{-i\epsilon_2})\,.
\eea

The refined topological vertex is defined as \cite{Iqbal:2007ii}:
\bea\label{TVdefinition}
&&C_{\lambda\mu\nu}(t,q)=\Big(\frac{q}{t}\Big)^{\frac{||\mu||^2}{2}}\,t^{\frac{\kappa(\mu)}{2}}\,q^{\frac{||\nu||^2}{2}}\,\widetilde{Z}_{\nu}(t,q)\,\\\nn
&&\times\sum_{\eta}\Big(\frac{q}{t}\Big)^{\frac{|\eta|+|\lambda|-|\mu|}{2}}\,s_{\lambda^t/\eta}(t^{-\rho}q^{-\nu})\,s_{\mu/\eta}(t^{-\nu^t}q^{-\rho})\,,
\eea
where $s_{\lambda/\mu}({\bf x})$ is the skew-Schur function and
\bea
\widetilde{Z}_{\nu}(t,q)=\prod_{(i,j)\in \nu}\Big(1-q^{\nu_{i}-j}\,t^{\nu^{t}_{j}-i}\Big)^{-1}\,.
\eea
For $q=t$ it reduces to the usual topological vertex \cite{Aganagic:2003db}.

\bea\label{ID2}
\sum_{\eta}s_{\eta/\lambda}(\mathbf{x})s_{\eta/\mu}(\mathbf{y})&=&\prod_{i,j=1}^{\infty}(1-x_{i}y_{j})^{-1}\sum_{\tau}s_{\mu/\tau}(\mathbf{x})s_{\lambda/\tau}(\mathbf{y})\,.\\
\sum_{\eta}s_{\eta^{t}/\lambda}(\mathbf{x})s_{\eta/\mu}(\mathbf{y})&=&\prod_{i,j=1}^{\infty}(1+x_{i}y_{j})\sum_{\tau}s_{\mu^{t}/\tau}(\mathbf{x})s_{\lambda^{t}/\tau^{t}}(\mathbf{y})\,.
\eea

\end{document}